\documentclass[conference,a4paper]{IEEEtran}
\IEEEoverridecommandlockouts

\usepackage[dvips]{graphicx}
\usepackage{url}
\usepackage{amsmath}
\usepackage{subfigure}
\usepackage{fancyhdr}

\begin{document}
\title{The Recommendation System to SNS Community for Tourists by Using Altruistic Behaviors
\thanks{\copyright 2016 IEEE. Personal use of this material is permitted. Permission from IEEE must be obtained for all other uses, in any current or future media, including reprinting/republishing this material for advertising or promotional purposes, creating new collective works, for resale or redistribution to servers or lists, or reuse of any copyrighted component of this work in other works.}
}

\author{\IEEEauthorblockN{Takumi Ichimura}
\IEEEauthorblockA{Dept. of Management and Systems,\\
Prefectural University of Hiroshima\\
\\
1-1-71, Ujina-Higashi, Minami-ku, \\
Hiroshima 734-8558, Japan\\
E-mail: ichimura@pu-hiroshima.ac.jp}
\and
\IEEEauthorblockN{Takuya Uemoto}
\IEEEauthorblockA{Graduated from\\
Graduate School of Comp. Sci. Research,\\
Prefectural University of Hiroshima.\\
1-1-71, Ujina-Higashi, Minami-ku, \\
Hiroshima 734-8558, Japan\\
E-mail: yslius7221@gmail.com}
\and
\IEEEauthorblockN{Shin Kamada}
\IEEEauthorblockA{Dept. of Intelligent Systems\\
Graduate School of Information Sciences\\
Hiroshima City University\\
3-4-1, Ozuka-Higashi, Asa-Minami-ku\\
Hiroshima, 731-3194, Japan\\
da65002@e.hiroshima-cu.ac.jp}
}

\maketitle

\pagestyle{fancy}{
\fancyhf{}
\fancyfoot[R]{}}
\renewcommand{\headrulewidth}{0pt}
\renewcommand{\footrulewidth}{0pt}

\begin{abstract}
We have already developed the recommendation system of sightseeing information on SNS by using smartphone based user participatory sensing system. The system can post the attractive information for tourists to the specified Facebook page by our developed smartphone application. The users in Facebook, who are interested in sightseeing, can come flocking through information space from far and near. However, the activities in the community on SNS are only supported by the specified people called a hub. We proposed the method of vitalization of tourist behaviors to give a stimulus to the people. We developed the simulation system for multi agent system with altruistic behaviors inspired by the Army Ants. The army ant takes feeding action with altruistic behaviors to suppress selfish behavior to a common object used by a plurality of users in common. In this paper, we introduced the altruism behavior determined by some simulation to vitalize the SNS community. The efficiency of the revitalization process of the community was investigated by some experimental simulation results.
\end{abstract}


\IEEEpeerreviewmaketitle

\section{Introduction}
A social community is composed of a group of individuals with the common interest and purpose. The size of community is determined by the number of people and their activities. The attraction to the community increases according to the integration of the interaction of the personal behaviors, and then the active behaviors grows to be a larger society. If we can find something which attracts us, we will use them to realize the extension of community. However, nobody knows what thing attracts interest to the people in the community.

We proposed the collaborative group developing method by mimicking altruism behavior of army ant agent system\cite{Ichimura15, Ichimura16}. The recommendation system of sightseeing information on social network service (SNS) is developed to the vitalize tourist behaviors in the community of SNS. The system is a kind of simulations for multi agent system with altruistic behaviors inspired by the Army Ants\cite{Ichimura14a, Ichimura14b}. The army ant takes feeding action with altruistic behaviors to suppress selfish behavior to a common object used by a plurality of users in common. In this paper, we introduced the altruism behavior from Army Ant simulation to vitalize the SNS community. The efficiency of the revitalization process of the community was investigated by some experimental simulation results. The experimental simulation analyzed the tourist sightseeing data collected by our developed smartphone application\cite{AndroidMarket}.

The current information technology can collect various data sets because the recent tremendous technical advances in processing power, storage capacity and network connected cloud computing. The sample record in such data set includes not only numerical values but also language, evaluation, and binary data such as pictures. Smartphone (Mobile) based Participatory Sensing (MPPS) systems involve a community of users sending personal information and participating in autonomous sensing through their mobile phones \cite{Lane2010}. Sensed data can be obtained from sensing devices on mobiles such as audio, video, and motion sensors by using smartphone application. Participation of smartphone users in sensorial data collection both from the individual and from the surrounding environment presents a wide range of opportunities for truly pervasive applications. The technical method to discover knowledge in such databases is known to be a field of data mining and developed in various research fields. The acquired knowledge works as filtering rules to select the submitted data from smartphone application to the SNS community.

We developed the smartphone application called ''Etajima Tourist Map\cite{AndroidMarket}'' to collect the information for attractive sightseeing spots and to acquire the filtering rules of finding new novelty spots. The community in the simulation system is the sightseeing information of Facebook page that shows the selected subjective data of the tourists as shown in Fig.\ref{fig:posting}. In order to verify the effectiveness of the proposed method to vitalize the community of tourists, the verification test was implemented. In this paper, we explain some experimental results.

\section{SmartPhone based Participatory Sensing System}
\label{sec:spps}
\subsection{Smartphone Application 'Etajima tourist Map'}
\begin{figure}[tbp]
\begin{center}
\subfigure[Start Display]{
\includegraphics[scale=0.1]{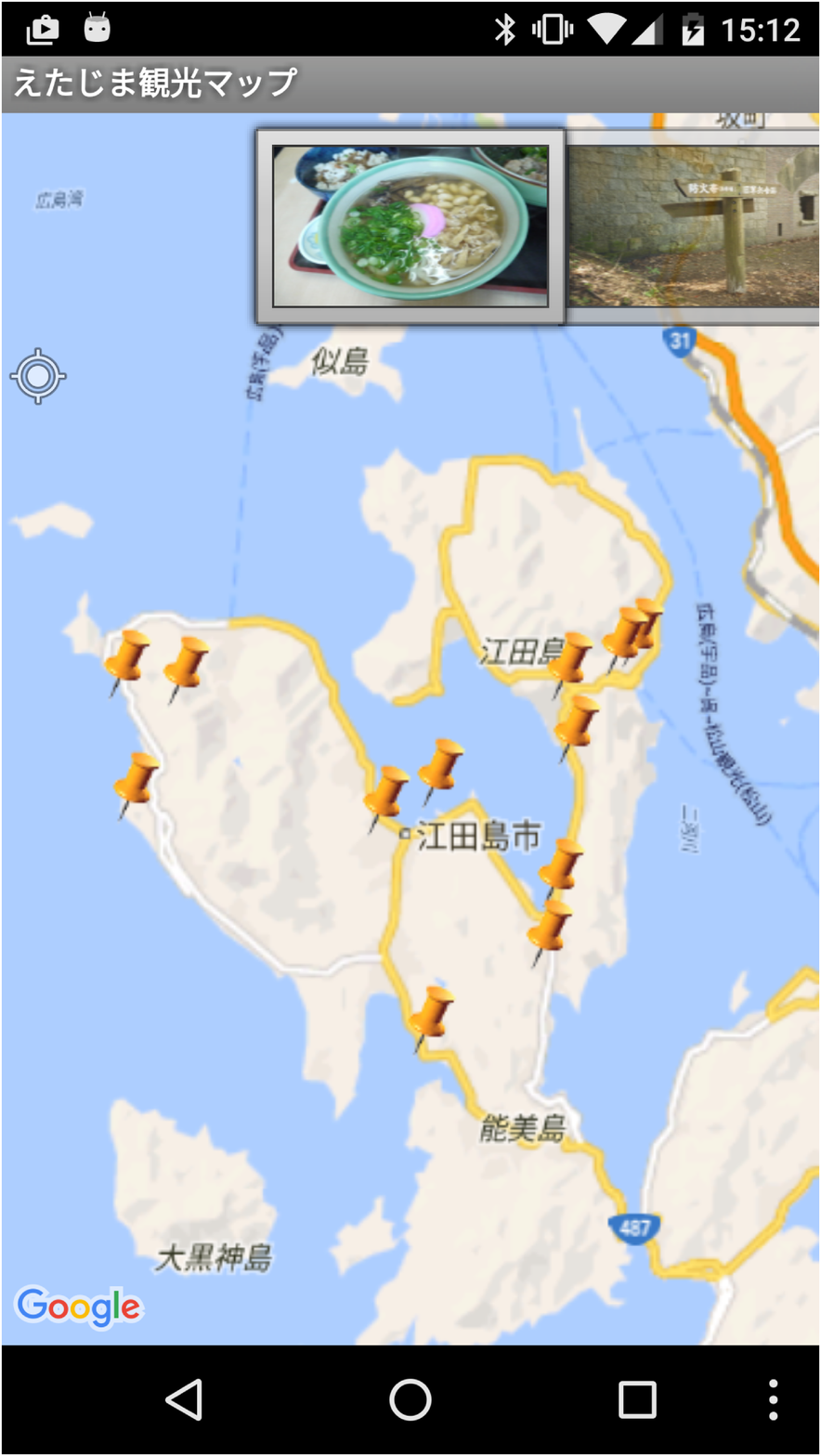}
\label{fig:KankouMap_Start}
}
\subfigure[List]{
\includegraphics[scale=0.1]{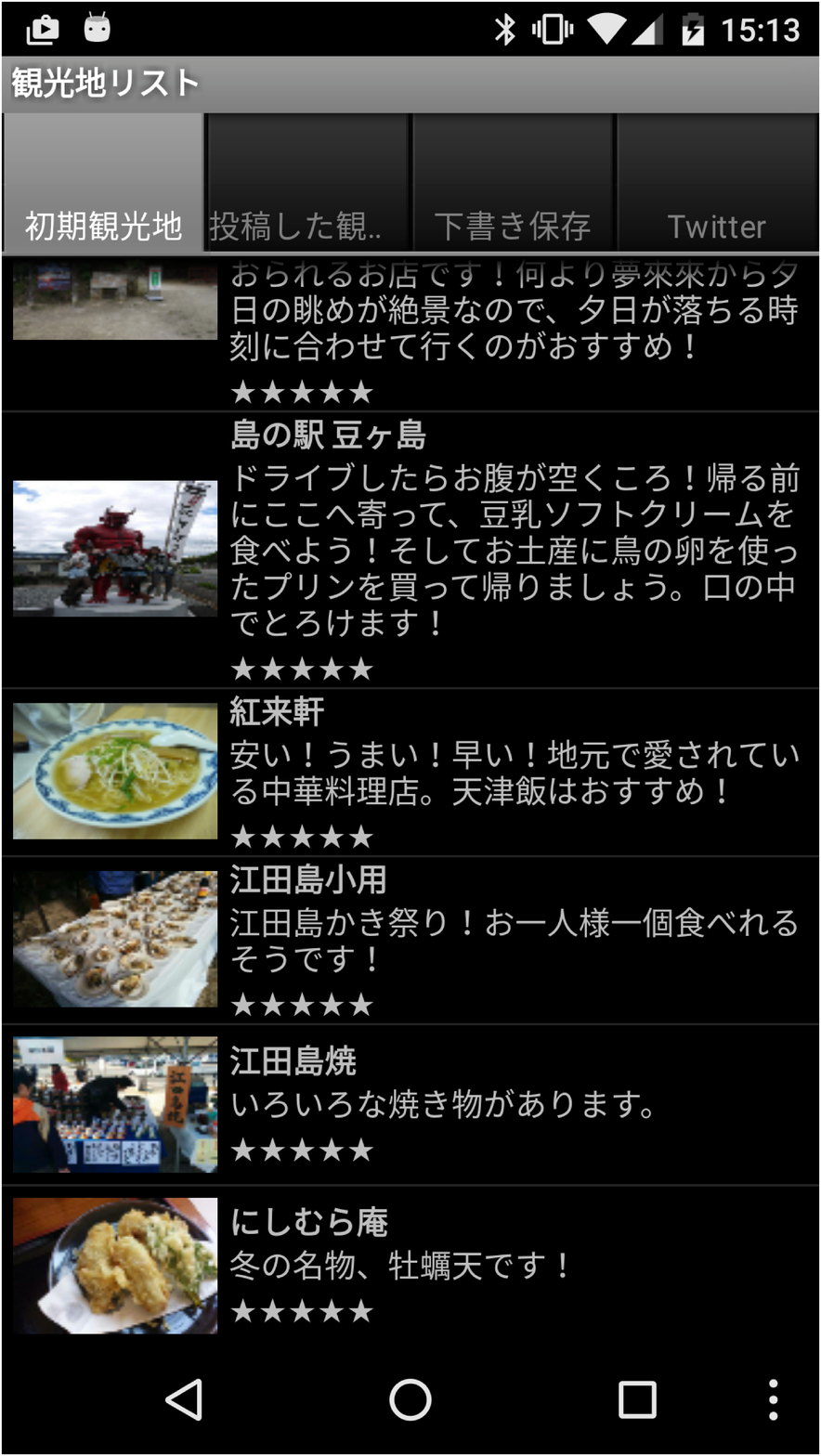}
\label{fig:KankouMap_List}
}
\subfigure[Posting Screen]{
\includegraphics[scale=0.125]{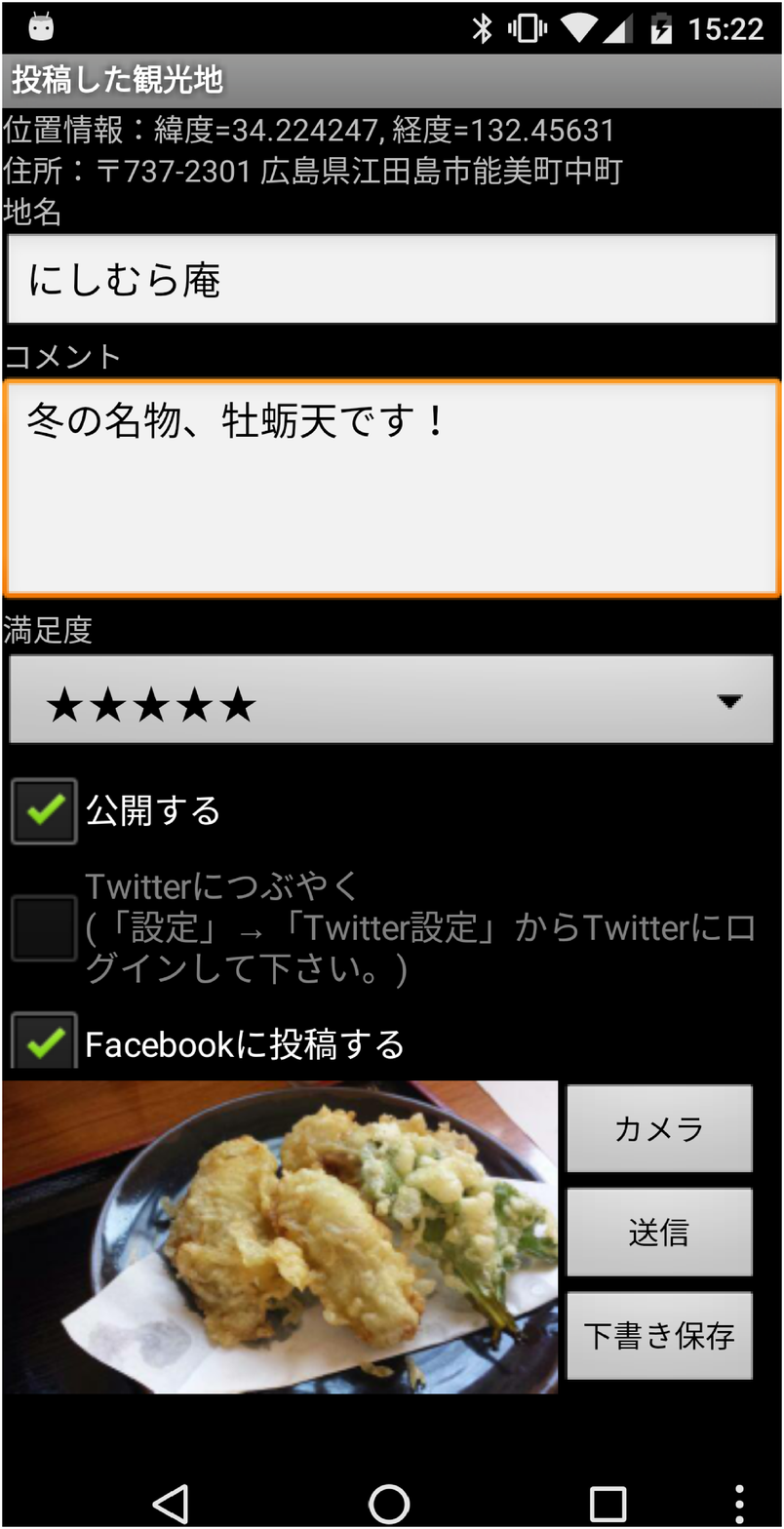}
\label{fig:KankouMap_Posting}
}
\subfigure[Posted Screen]{
\includegraphics[scale=0.1]{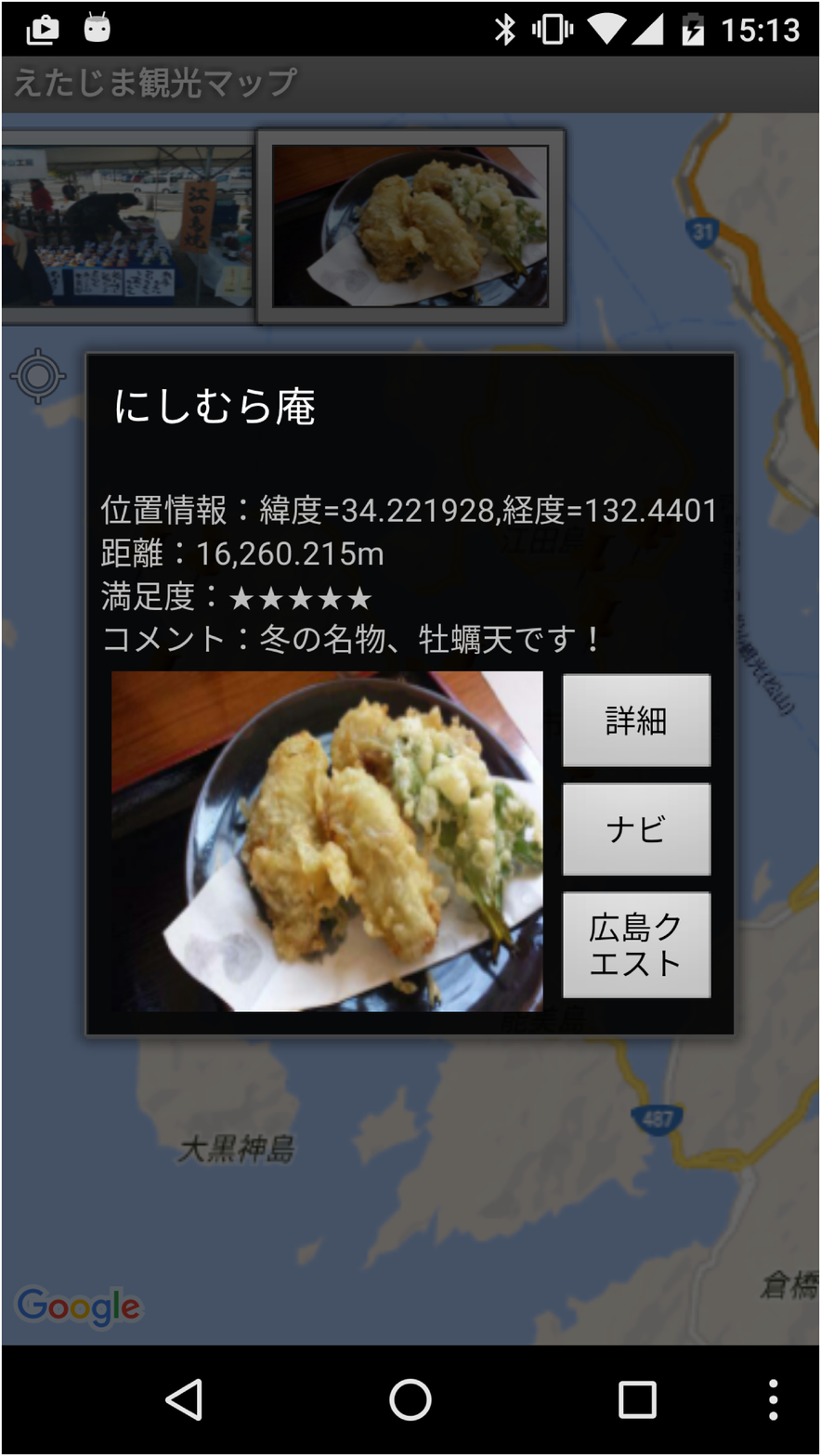}
\label{fig:KankouMap_Posted}
}
\caption{Etajima Tourist Map}
\label{fig:KankouMap}
\end{center}
\end{figure}

Participation of smartphone users in sensorial data collection both from the individual and from the surrounding environment presents a wide range opportunities of collecting firsthand information. For example, our MPPS developed smartphone application can collect the tourist subjective data as shown in Fig.\ref{fig:KankouMap}. The collected subjective data consist of jpeg files with GPS, geographic location name, the evaluation of $\{0, 1, 2, 3, 4\}$ and comments written in natural language at sightseeing spots to which a user really visits. The application must obtain GPS data before taking a picture so that the pictures provide evidence to prove that the tourist visited there.

The TF-IDF \cite{TF_IDF} is well known to calculate a weight used in information retrieval and text mining. The term frequency $tf(t,d)$ gives a measure of the importance of the term $t$ within the particular document $d$. The inverse document frequency $idf(t)$ is a measure of the general importance of the term. A high weight in $tfidf$ is reached by a high term frequency and a low document frequency of the term in the whole collection of documents. In this paper, $tf(t,d)$ is calculated from the comments to the website of tourist association.
 
\begin{eqnarray}
\nonumber tf(t, d)&=&\frac{n(t,d)}{\sum_{k} n(k,d)},\\
\nonumber idf(t) &=& \log \frac{|D|}{|\{d \in D: t \in d\}|},\\
tfidf(t,d)&=&tf(t,d) \times idf(t),
\label{eq:tfidf}
\end{eqnarray}
where $n(t,d)$ is the occurrence count of a term $t$ in the document $d$. $\sum_{k} n(k,d)$ is the occurrence count of all items in the document $d$. $|D|$ is the total number of documents in the corpus. $|\{d \in D: t \in d\}|$ is the number of documents where the term $t$ appears.

\subsection{An interface of interactive GHSOM}
\begin{figure}[htbp]
\begin{center}
\includegraphics[scale=0.4]{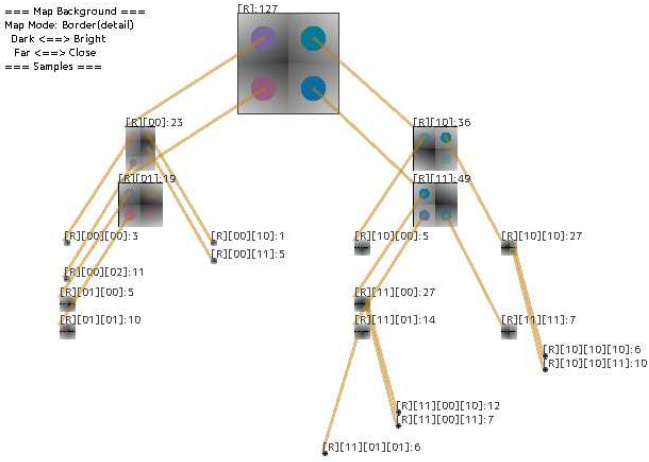}
\caption{Classification Result by Interactive GHSOM}
\label{fig:Etajima_InteractiveGHSOM}
\end{center}
\end{figure}

We developed the smartphone based interface of interactive GHSOM (Growing Hierarchical Self-Organizing Map) \cite{Ichimura11} to acquire the knowledge intuitively. This tool was developed by Java language. Fig.\ref{fig:Etajima_InteractiveGHSOM} shows the clustering results of the collected data by GHSOM. The notation $[R][01][01]:10$ in Fig.\ref{fig:Etajima_InteractiveGHSOM} means a category and represents the location of unit in the connection from the top level $[R]$. $[R]$ means a root node. [R][01] is the notation of subsequent lower layer ( away from root node). The notation can give suggestion of intuitive interpretation of classification of data. The numerical value('10') shows the number of samples divided into the leaf map after the sequence of classification $[R][01][10]$. The numerical values in the brackets mean the position of units in the corresponded map. The first letter (eg. `0') and the second letter (eg. `1') are the position in the column and the row in the map (eg. `01'), respectively.

The color of unit shows the pattern of sample represented in Munsell color system \cite{Munsell}, which consists of three independent dimensions: hue, value, and chroma. A color circle is an abstract illustrative organization of color hues around a circle. The similar color of units represents an intuitive understanding of similar pattern of samples in GHSOM.

\begin{figure}[htbp]
\begin{center}
\includegraphics[scale=0.35]{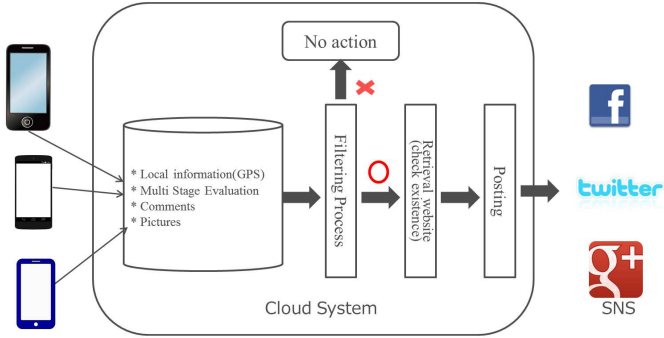}
\caption{Automatically Posting System to SNS }
\label{fig:posting}
\end{center}
\end{figure}

\subsection{Filtering Rules}
\begin{figure}[htbp]
\begin{center}
\includegraphics[scale=0.8]{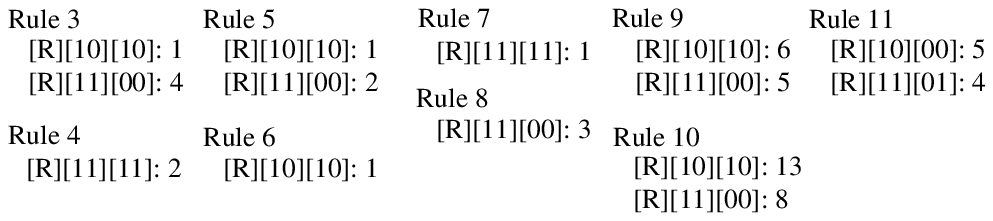}
\caption{Classification rules}
\label{fig:Filtering_rules}
\end{center}
\end{figure}

This subsection describes the filtering rules of tourist subjective data to be tweeted. Fig. \ref{fig:Filtering_rules} shows the overview of the clustering results by the interactive GHSOM and the decision tree by C4.5 via our developed smartphone application. The filtering rules are generated according to the acquired knowledge.

If the message to be tweeted matches the rule, the corresponding tourist information is important and the message is outgoing on social networking service and microblogging service as shown in Fig.\ref{fig:posting}.

In this paper, we have 11 filtering rules, the 9 pre-determined rules obtained from the collected data and 2 new rules in the proving test. Fig. \ref{fig:Filtering_rules} shows the only 9 filtering rules only, because Rule 1 and Rule 2 was not fired in this experiment.

\subsection{Relation between rules and activities in Facebook}
We investigated the relation between the information posted by the filtering rules and the degree of `Like' to its information on Facebook page by association rules\cite{Agrawal}. In this paper, the lift value was calculated to measure the quotient of the posterior and the prior confidence of association rule. In general, the lift value of the rule $R (=X \rightarrow Y)$ is as follows. 
\begin{eqnarray}
lift(R)&=&\frac{conf(X \rightarrow Y)}{conf(\phi \rightarrow Y)} = \frac{\frac{supp( X \cup Y)}{supp(X)}}{\frac{supp(Y)}{supp(\phi)}},\\
\label{eq:liftvalue}\\
conf( X \Rightarrow Y)&=& \frac{supp(X \cup Y)}{supp(X)},\\
\label{eq:Confidencevalue}
\end{eqnarray}
where $supp(\phi) = |T|$. $|T|$ is the number of the transaction.

\begin{figure}[htbp]
\begin{center}
\includegraphics[scale=0.7]{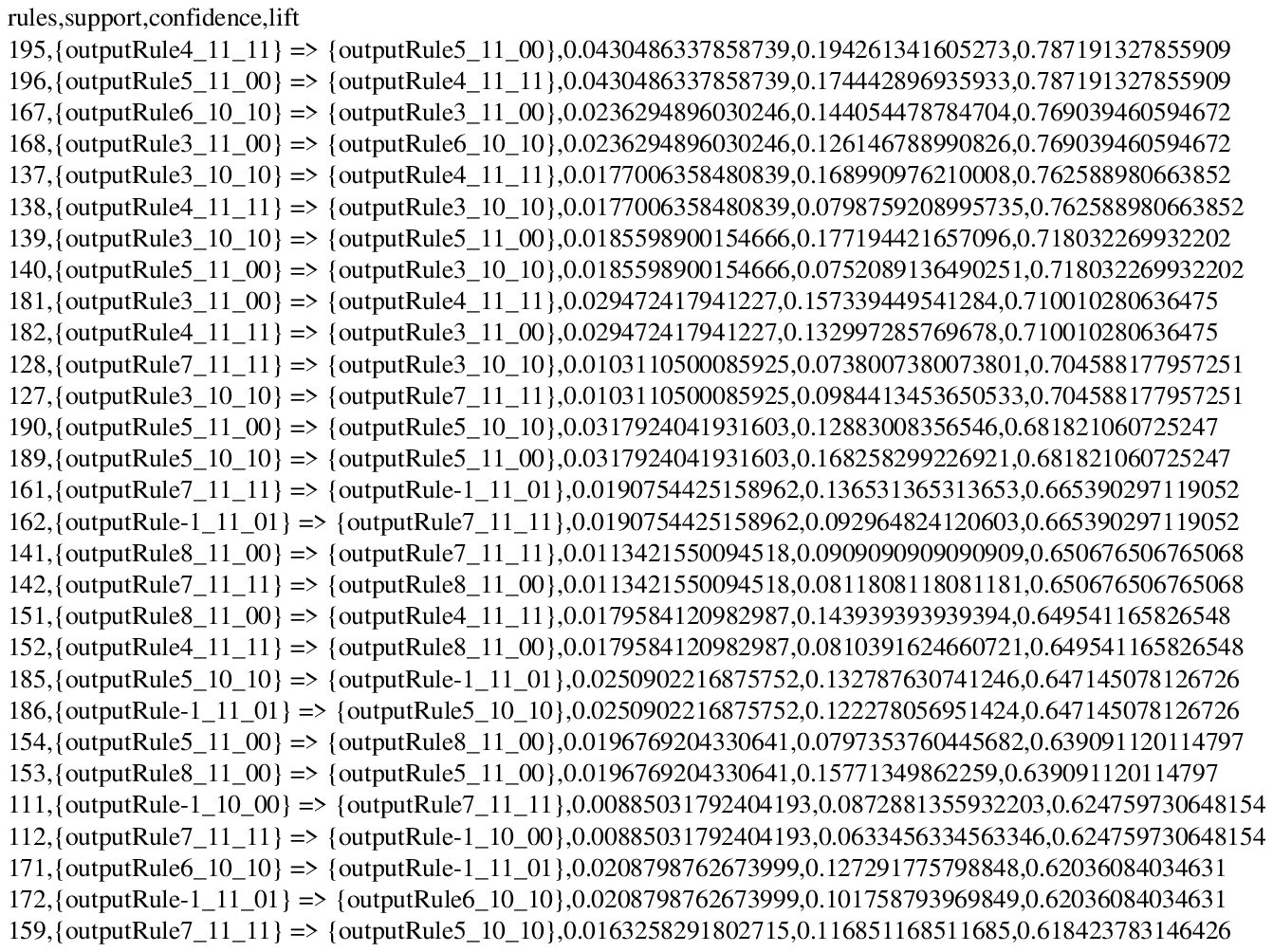}
\caption{Classification rules}
\label{fig:association_rules}
\end{center}
\end{figure}

First, select the user who pushes `Like' button to more than half of articles posted by filtering rules and categories. Second, check if the user pushes `Like' button to half of articles posted by other rules and categories. For example, the user, who pushes `Like' to more than half of articles posted by the category $[R][10][10]$ in Rule $5$ selects more than half articles posted by the category $[R][11][10]$ in Rule $8$. We investigated all user's behaviors and analyzed the relations by association rules. Fig.\ref{fig:association_rules} shows the example of calculation result which consists of the number of rules, `association rule', `support', `confidence', and `lift'. The higher lift value of 12 association rules are the combination of 4 rules: `Rule$9$,$[10][10]$', `Rule$9$,$[11][10]$', `Rule$10$,$[10][10]$', `Rule$10$,$[11][10]$' and the other higher lift value is 2 rules between `Rule$5$,$[10][10]$' and `Rule$6$,$[10][10]$'. For example, Rule $9$ and Rule $10$ among the filtering rules are related to the famous sightseeing spots in Etajima. From such observations, we developed the recommendation system of sightseeing spot with altruistic behaviors.

\section{Agent System with Altruistic Behavior}
\subsection{Altruistic Behavior in Army Ant Systems}
In animal societies, self-organization is the theory of how minimal complexity in the individual can generate greater complexity at the population. The rules specifying the interactions among the components are implemented by using only local information. In the field of social evolution, army ant performs altruism as one behavior of complexities, where each individual reduces its own fitness but increases the fitness of other individuals in the population. Such behaviors seem to be involved acts of self-sacrifice in order to aid the others. The concept was initially developed to explain the evolution of cooperation as mutually altruistic acts\cite{Trivers71}. The basic idea is close to the strategy of ``equivalent relation'' in the study of strategic decision making.

Army ants build a living nest with their bodies instead of building a nest like other ants. Each ant will hold on to the other legs and form a linked chain or a ball structure. This behavior is known as a bivouac. This allows the bridging of an empty space. 

We developed the multi-agent simulation system to execute more realistic altruistic behavior where two kinds of agents realize the division of roles in army ants\cite{Ichimura15}. Such observations of behaviors in the computer simulation will contribute to the shift to knowledge from the individual to the collective. We explored how agent-based modeling and simulation can be used as a research technique to study collaborative social networks\cite{Ichimura14a,Ichimura14b}. The situation for the forming the altruistic behavior becomes clear gradually in our research \cite{Ichimura15, Ichimura16}.

In our developed simulation system, there are 2 kinds of agents: `major agent' and `minor agent'. The major agent with 11 categories as shown in Fig.\ref{fig:Filtering_rules} works to post the article to Facebook and the minor agent views the posted one and push `Like' button to it by the agent's preference. The community of minor agents is constructed by the selection of the article and then we can find their agent with similar parameters to the preference of selecting articles. The articles posted by the specified major agent attracts some minor agents by the analyzed association rules.

The minor agent views and selects the preferred article by the following evaluation Eq.(\ref{eq:EtajimaLike}).
\begin{equation}
E^{k}_{i} \times S^{k}_{j} \geq L_{threshold}, 
\label{eq:EtajimaLike}
\end{equation}
where the evaluation value $E^k_i$ represents the attractive degree to the article $i$ in the category $k$. The $S^k_j$ represents the interesting degree to the category $k$ by the minor agent $j$. The $S^k_j$ was given to the agent when it is generated. The higher value of $S^k_j$ means the agent strongly interested in the category. $L_{threshold}$ is a threshold value for the selection.

\subsection{Altruistic Behavior in Simulation model}
The altruistic behaviors seem to be involved acts of self-sacrifice in order to aid the others. In this model, altruistic behavior means the recommendation of the article posted by the specified minor agent to the other agents. In the SNS community such as Facebook, the behavior of the recommendation is sharing the articles to the time line. The minor agents decides to take the following altruistic behavior, Eq.(\ref{eq:Etajimaalt}) with a certain probability $P(Alt)$.

\begin{eqnarray}
\left\{
\begin{array}{l}
If\: E^{k}_{i} \geq 3 \:and \: R_{im} \times \frac{E^{k}_{i} \times S^{k}_{j}}{N_{i}} \geq A_{threshold}\\
\qquad Then \: AltruisticBehavior\\
Otherwise\;\; Then \; NotAltruism
\end{array}
\right.
\label{eq:Etajimaalt}
\end{eqnarray}

$R_{im}=1$ is the condition that the article was posted by the specified major agent. $E^{k}_{i}$ is the evaluation value in the category $k$ and $S^{k}_{l}$ is the interesting degree for the category $k$ by the minor agent $l$. $N_{i}$ is the number of smartphone users who push `Like' button to the articles $i$. $A_{threshold}$ is the threshold value. $E^{k}_{i} \times S^{k}_{j}$ calculates the attractive degree by the posting. The higher value of $E^{k}_{i} \times S^{k}_{j}$ recommend the article. If $N_{i}$ is large, the altruistic behavior decreases, because $E^{k}_{i} \times S^{k}_{j}$ becomes smaller.

The minor agent $j$ with altruistic behavior selects an article from what the agent pushes `Like' button. The agent recommends the other agents and then they push `Like' to the same article. The agent decides the selection of the article by Eq.(\ref{eq:EtajimaLike}).

\section{Experimental Result for Community Vitalization}
\subsection{Optimal value for threshold}
\begin{figure}[h]
\begin{center}
\includegraphics[scale=0.7]{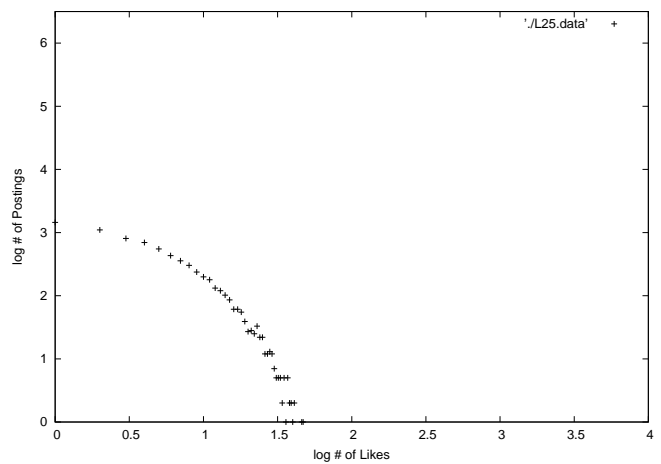}
\caption{Simulation Result for $L_{threshold}$($=2.5$)}
\label{fig:L_threshold-4false}
\end{center}
\end{figure}

We investigated the optimal set for 3 parameters $L_{threshold}$, $A_{threshold}$, $P(Alt)$ by some experimental results. Each experiment has 200 agents, and 2,000 minor agents for 100 steps. First, Fig.\ref{fig:L_threshold-4false} shows the relation between $L_{threshold}$ and the activities in the community. We investigated the experiments with respect to $L_{threshold}=\{0.5, 1.5, 2.5, 3.5, 4.5\}$. X axis is the logarithm of the number of `Like' by Eq.(\ref{eq:EtajimaLike}) and T axis is the logarithm of the number of articles. There are many articles with less `Like' as a matter of course. As the $L_{threshold}$ is higher, the preferred article get fewer. That is, the article selection criterion by the minor agents  is strict in case of high $L_{threshold}$ and the selected article is high evaluation by many agents. Fig.\ref{fig:L_threshold-4false} shows the case of 
$L_{threshold}=2.5$.

\begin{figure}[h]
\begin{center}
\includegraphics[scale=0.7]{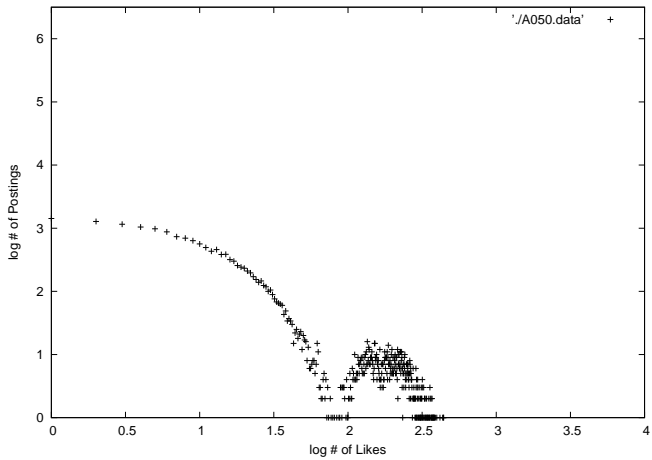}
\caption{Simulation Result with Altruistic Behavior ($A_{threshold}=0.05$)}
\label{fig:A_threshold_1}
\end{center}
\end{figure}

Second, we implemented some examinations related to $A_{threshold}=\{0.05, 0.075, 0.10, 0.125, 0.15\}$ with $L_{threshold}=2.5$. X axis and Y axis indicate the same one in the experiments of $L_{threshold}$, respectively. The $A_{threshold}$ occurs that many minor agents view the articles. If the article is good, the number of `Like' is large. The phenomenon is caused by the effects for the altruistic behaviors. The results for all $A_{threshold}$ were divided into 2 kinds of distributions as shown in Fig.\ref{fig:A_threshold_1}. The latter distribution was formed by the altruistic behaviors. We can observe that the community grows largely based on the recommendation by many agents with small $A_{threshold}$. The case for $A_{threshold}=0.05$ was best situation.

\begin{figure}[h]
\begin{center}
\includegraphics[scale=0.2]{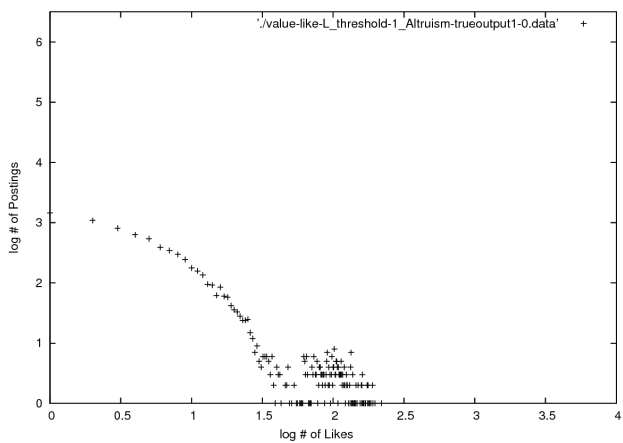}
\caption{Simulation Result with Altruistic Behavior ($L_{threshold}=2.5$, $A_{threshold}=0.05$, $P(Alt)=1.0$)}
\label{fig:P_Alt_10}
\end{center}
\end{figure}

Third, the experiments for parameter $P(Alt) (=\{0.2,0.4,0.6,0.8,1.0\})$ was executed by using results of $L_{threshold}=2.5$, $A_{threshold}=0.05$. Fig.\ref{fig:P_Alt_10} shows the case of $P(Alt)=1.0$. As for the variation of $P(Alt)$, there are no difference among experimental results. 

\begin{table*}[htbp]
\begin{center}
\caption{Simulation Result with Altruistic Behaviors}	
\label{tab:result2}
\begin{tabular}{|c|c|c|c|c|c|c|c|c|c|}
\hline
	&	\multicolumn{4}{c}{without Altruistic} 	&	\multicolumn{4}{|c|}{with Altruistic}		&	\\\cline{2-9}
Category       	&	Articles Ave.&	Like Ave.     	&	 Num of Person 	&	ratio 	&	Articles Ave.&	Like Ave.     	&	 Num of Person 	&	ratio & Diff.\\	\hline
3	&	1365	&	253.54	&	789	&	0.3213	&	1363	&	270.82	&	789	&	0.3432	&	17.28	\\	\hline
4	&	546	&	261.34	&	816	&	0.3203	&	544	&	277.37	&	816	&	0.3399	&	16.03	\\	\hline
5	&	819	&	253.57	&	792	&	0.3202	&	816	&	270.37	&	792	&	0.3414	&	16.80	\\	\hline
6	&	273	&	251.07	&	780	&	0.3219	&	272	&	264.09	&	780	&	0.3322	&	13.02	\\	\hline
7	&	273	&	269.74	&	845	&	0.3192	&	272	&	284.81	&	845	&	0.337	&	15.07	\\	\hline
8	&	819	&	256.67	&	801	&	0.3204	&	816	&	272.87	&	801	&	0.3407	&	16.20	\\	\hline
9	&	3003	&	245.69	&	768	&	0.3199	&	2992	&	262.63	&	768	&	0.342	&	16.94	\\	\hline
10	&	5458	&	258.62	&	811	&	0.3189	&	5440	&	277.76	&	811	&	0.3425	&	19.14	\\	\hline
11	&	2448	&	155.76	&	488	&	0.3192	&	2448	&	167.43	&	488	&	0.3431	&	11.67	\\	\hline\hline
Average     	&	1667.11	&	245.11	&	765.56	&	0.3201	&	1662.56	&	260.35	&	765.56	&	0.3402	&	15.79	\\	\hline
Variance     	&	1613.8	&	32.23	&	100.38	&	0.0009	&	1608.83	&	33.66	&	100.38	&	0.0034	&	4.57	\\	\hline
\end{tabular}
\end{center}
\end{table*}

\begin{figure}[htbp]
\begin{center}
\includegraphics[scale=0.5]{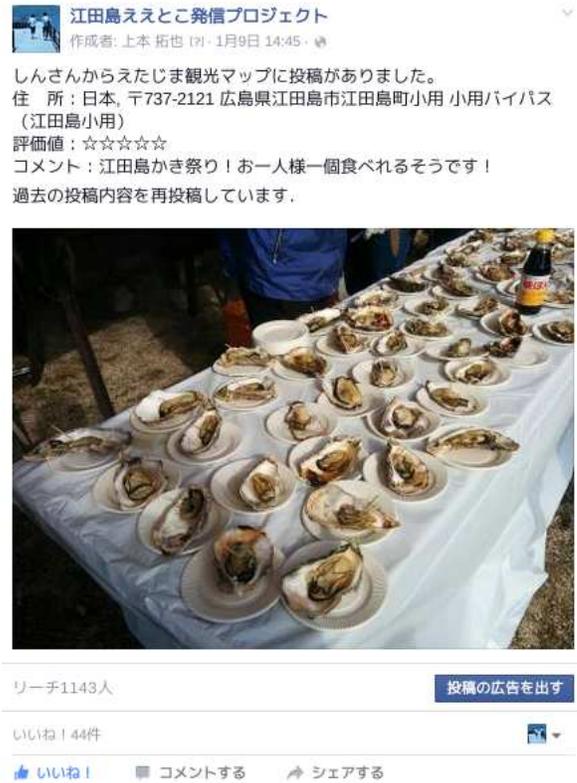}
\caption{The article by automatically posted by MPPS}
\label{fig:Etajima_Reposting}
\end{center}
\end{figure}

We applied the simulation model with $L_{threshold}=2.5$, $A_{threshold}=0.05$, $P(Alt)=1.0$ to the smartphone based participatory system described in Section \ref{sec:spps}. The posted article is represented on Facebook page as shown in Fig.\ref{fig:Etajima_Reposting}.

Table \ref{tab:result2} shows the simulation result with altruistic behaviors to our developed Etajima sightseeing system. The `Category' is the classified by Interactive GHSOM. The `Articles Ave.' and the `Like Ave.' are the average of number of articles and `Like', respectively. The `Num of Person' is the number of persons who pushes `Like' to the articles in the category. The `ratio' is the articles divided by the number of persons. The `Diff.' means the difference of the number of `Like'. As shown in Table \ref{tab:result2}, the number of `Like' becomes larger by introducing the altruistic behaviors.

Table \ref{tab:result3} shows the simulation results with the re-posting articles based on the system's recommendation instead of users. The articles were recommended by the system. Every numerical values in Table \ref{tab:result3} are small, because the original data were collected within the limited periodical time. As for `A', `B', and `G', the number of `Like' becomes large and the number of `reach' increases drastically. The `reach' means that the article is spread to their community. As for `D' and `E', the number of `reach' increases. As for `G', the simulation test executed in winter but the article is related to summer. In case of `C' and `F', we can see an interesting feature of the decrease of `Likes'. These articles were very famous tourist attractions such as `oysters'. The SNS users are fed up too famous articles, because they see the similar contents in their timeline on SNS. Therefore, re-posting their articles is not very attractive to users. On the other hand, when the friends of the SNS user see the famous articles at first time, they feel it so attractive. As a result, the number of `reach' of `C' and `F' increases.

\begin{table}[hbp]
\begin{center}
\caption{Re posting for Community Vitalization}
\label{tab:result3}
\begin{tabular}{|c|c|c|c|c|}
\hline
	&	\multicolumn{2}{c|}{Smartphone}	&	\multicolumn{2}{c|}{Reposting}	\\\hline
Information		&	Likes	&	Reach	&	Likes	&	Reach	\\\hline
A 	&	60	&	435	&	73	&	1837	\\\hline
B	&	43	&	223	&	66	&	1518	\\\hline
C		&	81	&	1063	&	45	&	1220	\\\hline
D	&	44	&	560	&	50	&	1322	\\\hline
E	&	24	&	170	&	31	&	1141	\\\hline
F	&	64	&	390	&	34	&	1074	\\\hline
G		&	27	&	165	&	41	&	1124	\\\hline
H		&	52	&	645	&	37	&	987	\\\hline
\end{tabular}
\end{center}
\end{table}

\section{Conclusion}
We have already developed the recommendation system of sightseeing information on SNS by using smartphone based user participatory sensing system. The system can post the attractive information for tourists to the specified Facebook page by our developed smartphone application. To vitalize the tourist behaviors in the community on SNS, we examined simulation tests for agent behaviors with altruism. The altruistic behavior is inspired by the Army Ants and the simulation system for multi agent system with altruistic behaviors has developed. In this paper, we introduced the altruism behavior of army ant simulation system to vitalize the SNS community. The efficiency of the revitalization process of the community was investigated by some experimental simulation results. Such total system is useful to activate the local community such as providing information system to tourist.

\section*{Acknowledgment}
This work was supported in part by the JSPS KAKENHI Grant Number 25330366.


\begin{thebibliography}{1}
\bibitem{Ichimura15}
T.Ichimura, and T.Uemoto,
`Analysis of the Social Community Based on the Network Growing Model in Open Source Software Community',
\emph{Proc. of IEEE 8th International Workshop on Computational Intelligence and Applications (IWCIA2015)}, pp.149-153(2015)

\bibitem{Ichimura16}
T.Ichimura, T.Uemoto, S.Kamada,
`A Collaborative Group Developing Method of Open Source Software by Mimicking Altruism Behavior of Army Ant',
\emph{Proc. of IMECS2016, Intl. MultiConf. of Engineering and Computer Scientists}, Vol.1, pp.46-51(2016)

\bibitem{Ichimura14a}
T.Ichimura, T.Uemoto, A.Hara,
`Emergence of Altruism Behavior for Multi Feeding Areas in Army Ant Social Evolutionary System',
\emph{Proc. of 2014 IEEE International Conference on Systems, Man, and Cybernetics (IEEE SMC 2014)}, pp.176-181 (2014)

\bibitem{Ichimura14b}
T.Ichimura, T.Uemoto, A.Hara, K.J.Mackin,
`Emergence of Altruism Behavior in Army Ant Based Social Evolutionary System',
\emph{SpringerPlus, A Springer Open Journal}, Vol.3, 712, doi:10.1186/2193-1801-3-712(2014).

\bibitem{Lane2010}
N.D.Lane, E.Miluzzo, L.Hong, D.Peebles, T.Choudhury, A.T.Campbell,
`A survey of mobile phone sensing', 
\emph{IEEE Communications Magazine},
Vol.48, No.9, pp.140-150 (2010)

\bibitem{AndroidMarket}
ITProducts,
`Etajima Hiroshima Tourist Map',
\url{https://play.google.com/store/apps/ details?id=jp.skproducts.KankouMap_Etajima&hl=ja}
Retrieved 2016-7-19(2016)

\bibitem{Rauber02}
A.Rauber, D.Merkl, M.Dittenbach, 
`The growing hierarchical self-organizing map: exploratory analysis of high-dimensional data',
\emph{IEEE Transactions on Neural Networks}, vol.13, pp.1331-1341(2002)

\bibitem{Quinlan96}
J.R.Quinlan,
`Improved use of continuous attributes in c4.5',
\emph{Journal of Artificial Intelligence Research},
No.4, pp.77-90(1996)

\bibitem{TF_IDF}
H.C.Wu, R.W.P.Ruk, K.F.Wong, K.L.Kwok,
`Interpreting TF-IDF term weights as making relevance decisions',
\emph{ACM Transactions on Information Systems},
Vol.26, No.3, pp.137 (2008)

\bibitem{Ichimura11}
T.Ichimura, T.Yamaguchi,
`A Proposal of Interactive Growing Hierarchical SOM',
Proc. of 2011 IEEE International Conference on Systems, Man, and Cybernetics (IEEE SMC 2011), pp.3149-3154 (2011)

\bibitem{Munsell}
R.G.Kuehni,
`The early development of the Munsell system',
\emph{Color Research and Application},
Vol.27, No.1, pp.2027 (2002).

\bibitem{Agrawal}
R.Agrawal, R.Srikant,
`Fast algorithms for mining association rules',
\emph{Proc. 20th Int. Conf. Very Large Data Bases}, VLDB, pp.487-499(1994)

\bibitem{Trivers71}
Trivers RL, 
\emph{The evolution of reciprocal altruism},
Q Rev Biol, Vol.46, pp.35-67(1971)

\bibitem{Ichimura12}
T.Ichimura, Y.Douzono Y,
`Altruism Simulation based on Pheromone Evaporation and Its Diffusion in Army Ant Inspired Social Evolutionary System',
\emph{Proc. The 6th International conference on Soft Computing and Intelligent Systems and The 13th International Symposium on Advanced Intelligent Systems (SCIS-ISIS2012)}
pp. 1357--{1362}

\end{thebibliography}
\end{document}